\documentclass{article}
\usepackage[utf8]{inputenc}
\usepackage{caption}
\usepackage{subcaption}
\usepackage{graphicx}
\usepackage{tikz}
\usepackage{graphicx,pgf}
\usetikzlibrary{decorations.pathreplacing}
\usetikzlibrary{arrows,shapes}
\usepackage{amsmath}
\usepackage{amsfonts}
\usepackage{amssymb}
\usepackage{hyperref}
\usepackage{fullpage}
\usepackage{cancel}
\usepackage{enumerate}
\usepackage{comment}
\usepackage{enumitem}
\usepackage{setspace}

\newcommand{\del}{\partial}
\newcommand{\grad}{\nabla}
\newcommand{\scri}{\mathcal{I}}
\newcommand{\scrip}{\mathcal{I}^+}
\newcommand\Lie{\mathcal{L}}
\newcommand{\Or}{\mathcal{O}}
\newcommand{\ord}{\mathcal{O}}

\newcommand{\eref}[1]{eq.\,(\ref{#1})}

\title{Power radiated by a binary system in a de Sitter Universe}

\author{B\'eatrice Bonga\footnote{Electronic address: bpb165@psu.edu} \\
  \multicolumn{1}{p{.7\textwidth}}{\centering\emph{Institute for Gravitation and the Cosmos \& Physics Department, The Pennsylvania State University}} \\
  \\  
  Jeffrey S. Hazboun\\
  \multicolumn{1}{p{.7\textwidth}}{\centering\emph{Center for Advanced Radio Astronomy, \\ University of Texas Rio Grande Valley}}
  }

\allowdisplaybreaks
\begin{document}
\maketitle

\begin{abstract}
Gravitational waves emitted by high redshift sources propagate through various epochs of the Universe including the current era of measurable, accelerated expansion.
Historically, the calculation of gravitational wave power on cosmological backgrounds is based on various simplifications, including a $1/r$-expansion and the use of an algebraic projection to retrieve the radiative degrees of freedom. On a de~Sitter spacetime, recent work has demonstrated that many of these calculational techniques and approximations do not apply. Here we calculate the power emitted by a binary system on a de~Sitter background using techniques tailored to de Sitter spacetime.
The common expression for the power radiated by this source in an FLRW spacetime, calculated using far wave-zone techniques, gives the same expression as the late time expansion specialized to the de~Sitter background in the high-frequency approximation. 
\end{abstract}

\section{Introduction}
With the recent detections by LIGO of a coalescing binary black hole system, the era of gravitational wave astronomy has begun \cite{Abbott:2016blz, Abbott:2016nmj,PhysRevLett.118.221101}. 
As ground-based detectors improve their sensitivity, space-based detectors start construction and pulsar timing arrays begin detecting the stochastic background of super massive blackholes, the era of gravitational wave cosmology will dawn. 
These efforts do not only require tremendous experimental efforts and intricate data analysis, but also a strong theoretical understanding of the relevant physics involved. 
The development of gravitational wave theory that was pivotal for the recent detection, started in the 1960s and is based on the framework of asymptotically flat spacetimes \cite{Bondi:1962, Sachs:1962}. 
However, from astrophysical and cosmological observations, we know that our Universe is best
described by the inclusion of a positive cosmological constant $\Lambda$  \cite{Perlmutter:1998np,Riess:1998cb,Ade:2013zuv}. 
Future gravitational wave observations - in particular those aimed at detecting gravitational waves at high redshifts - will require a better understanding of how gravitational waves are affected by the cosmological constant. 
One of the most powerful sources of gravitational waves is the interaction of two supermassive black holes following the merger of their two host galaxies. 
While the details of this evolution are still being investigated \cite{Ravi:2014aha,Taylor:2016ftv}, the event rates of coalescences in a given volume are small enough that we will only observe them at cosmological distances. 
Thus, for such physical systems it is essential to understand the effect of $\Lambda$.
Recent research has shown that regardless of how small the value of the cosmological constant is, many surprises and difficulties arise when $\Lambda$ is no longer exactly zero \cite{ABK1,ABK2,abk3,ABK4}.
This is illustrated by the discontinuous behavior of some physical observables, such as energy, in the limit $\Lambda \to 0$.

In this paper, our aim is to calculate the power radiated by a binary system on a de~Sitter background. 
This nicely compliments earlier work on the scalar and electromagnetic radiation emitted by charges on a de Sitter background \cite{Krtous:2003rw, Bicak:2005yt, Faci:2011rq, Blaga:2016dsu} and radiation from uniformly accelerated charged black holes with $\Lambda>0$ \cite{Krtous:2003tc}.
We restrict ourselves to a binary system whose orbital motion is well-approximated by a circular orbit.
A good understanding of the amount of power radiated by such a relatively simple system is critical for both indirect and direct observations of gravitational waves emitted by binaries. 
If the power radiated in the form of gravitational waves is altered by the cosmological constant, the shrinking of the orbit and the decay in period will be modified as well. 
This would affect indirect observations of gravitational waves that observe the time derivative of a binary's rotational period over long periods of time \cite{Hulse:1974eb,Taylor:1982zz},
as well as direct observations since the power directly influences the evolution of the waveform \cite{Peters:1963ux, hughesetal:2005}.
In addition, a thorough comprehension of the power radiated by this system is interesting from a theoretical standpoint. 
A general formula for gravitational power radiated by a source on a de Sitter background was recently discovered \cite{abk3}. 
This is the first application of that formula. It illustrates nicely some of the general properties of gravitational power in a concrete example as well as highlighting aspects unique to the power emitted by a binary system.

The current understanding of propagation effects on a cosmological background such as a Friedmann-Lema\^{i}tre-Robertson-Walker (FLRW) spacetime, relies heavily on techniques borrowed from ``far-away wave zone'' calculations in flat spacetimes \cite{thorne}.
In particular, the calculation of power radiated by a physical system on an FLRW spacetime uses the $1/r$ expansion, an algebraic projection operator to extract the radiative modes from the linearized gravitational perturbation, and some type of spatial or time average over one (or several) reduced wavelengths or complete orbital periods of the source.
Since de Sitter spacetime is globally very different from Minkowski spacetime, and most FLRW spacetimes, \emph{none of these calculational tools extend to de Sitter spacetime}. 
The $1/r$ expansion is to be replaced by a `late time' expansion. Instead of using the algebraic projection operator to extract the radiative modes, one needs to calculate the transverse-traceless part of the perturbation by solving a set of differential equations.
Moreover, no time averaging is done, as we work with a late time limit, nor is any spatial averaging needed.

Despite these drastic differences in calculational tools and their associated approximations, we show that the final result for power radiated by a binary system on a de Sitter background is  the same as for a binary in an FLRW spacetime in the high-frequency approximation.
Specifically, we find that the power radiated is redshift independent when expressed in terms of physical source quantities such as its reduced mass, orbital angular velocity and separation between the two bodies making up the binary.
This is the first calculation \emph{from first principles} that shows that the numerical effect of the cosmological constant $\Lambda$ on the power radiated by a binary is indeed small given the observed value of $\Lambda$. 
This does not rule out, however, that $\Lambda$-dependent corrections may be important for the phase shift in the gravitational waveform, which is an integrated effect of the power.

The paper is organized as follows. 
First, we review the general formula for power on a de Sitter background in Sec.\ref{sec:powerbackground}. 
Next, in Sec.~\ref{sec:physicalsetup}, we carefully spell out all the approximations made to model a binary on a de Sitter background.  
Given these approximations, the resulting power is presented in Sec.~\ref{sec:powerresult}. This is the heart of the paper.
Details of the calculation of the transverse-traceless part, needed to calculate the power, are included in the Appendix. 
Sec.~\ref{sec:discussion} contains a summary of the result.
We use the convention $c=1$ throughout.



\section{Preliminaries: Formula for power radiated when $\Lambda>0$}
\label{sec:powerbackground}
The physical system of interest is a binary system at a 
cosmological distance.
We model this system by a linearized source on a 
cosmological background spacetime. Ideally, one would like to consider a realistic 
FLRW model describing the different epochs 
of the 
Universe that the gravitational radiation emitted by this system would have traveled
through. Here we take the approach that we would like to know the \textit{maximal} possible
effect of the cosmological constant on the gravitational radiation. 
Therefore, we ignore the radiation- and matter-dominated epochs of our 
cosmological history and consider a de~Sitter background\footnote{A purely de~Sitter background is also appropriate for the stochastic background of gravitational waves from an inflationary period, but that radiation is source-free.}. 
We use the framework developed in \cite{abk3} to calculate the power emitted by this system. 
In this section, the relevant equations from \cite{abk3} are summarized
and discussed. Readers familiar with this work can safely skip this section.\\

The causal future of the binary system only covers the Poincar\'e patch of the de~Sitter spacetime. 
Therefore, to study the gravitational radiation emitted by this isolated system, it suffices to restrict ourselves to this patch. The metric in $(t,x,y,z)-$coordinates adapted to this patch is:
\begin{equation}
\label{eq:FLRWmetric}
g_{ab} = - \grad_a t \, \grad_b t + a^2(t) \left[\grad_a x \, \grad_b x + \grad_a y \, \grad_b y + \grad_a z \, \grad_b z \right]
\end{equation}
with the scale factor $a= e^{\sqrt{\frac{\Lambda}{3}} t}$. 
(For an overview of other coordinate systems of de Sitter spacetime, see the extensive appendix of \cite{Bicak:2005yt}.)
The `time translation' vector field of de~Sitter spacetime in these coordinates is  
\begin{equation}
T^a \partial_a = \frac{\partial}{\partial t}
- \sqrt{\frac{\Lambda}{3}} \left(x  \frac{\partial}{\partial x} 
+y \frac{\partial}{\partial y}
+z \frac{\partial}{\partial z} \right).
\end{equation}
The Hamiltonian associated with this Killing vector field defines the energy and the
Lie derivative of this Hamiltonian with respect to $T^a$ defines the power. Note that
this vector field is not globally timelike. Nevertheless, it is considered a time 
translation vector field for two reasons: 
(i) in the limit $\Lambda \to 0$, one recovers the time translation
vector field of Minkowski spacetime, and 
(ii) it is the limit of the time translation Killing vector
field in Schwarzschild-de~Sitter spacetimes as the Schwarzschild mass is taken to 
zero.

Analogously to the power radiated for perturbations on a Minkowski background, the 
power radiated by a source on a de~Sitter background is expressed in terms of 
its quadrupole moments. On a constant time slice, the Cartesian components of the 
quadrupole moment in de~Sitter spacetime are
\begin{eqnarray}
&Q_{ab}^{(\rho)} & (t) := \int \! dx \! \int \! dy \! \int \! dz  \, a^3(t) \; \rho(t) \, \bar{x}_a \, \bar{x}_b  \\
&Q_{ab}^{(p)} (t) & := \int \! dx \! \int \! dy \! \int \! dz  \, a^3(t) \; \left(p_x(t) + p_y(t) + p_z(t) \right) \, \bar{x}_a \, \bar{x}_b  
\end{eqnarray}
where $\bar{x}_a := a(t) x_a$ is the physical separation of a point described 
by the coordinates $\vec{x}$ to the origin, $\rho$ is the energy density 
and $p_i$ is the pressure in the $i-$direction. The mass quadrupole moment is indicated by the superscript $\rho$ and the pressure quadrupole moment by $p$. These quadrupole moments describe (part of) the physical attributes of the source and are defined for any constant time-slice.

In order to extract the physics of an isolated system unambiguously, 
the formula for power radiated is defined at null infinity\footnote{This 
is similar to what is done for isolated systems when $\Lambda=0$.
There one works with asymptotically flat spacetimes to study isolated systems and even though the detector 
is at a finite distance from the source, one generally performs calculations at 
$\scri^+$ -- which is at ``$r = \infty$''. 
This is  because the expressions of radiated energy and momentum are 
unambiguous at $\scri^+$. Similarly, for the case with $\Lambda>0$, the detector will 
not be at the infinite future of the source either, but following the same logic 
as for asymptotically flat spacetimes, we calculate the power at $\scri^+$.
},  $\scri^+$.
This surface is reached using a late time expansion, rather than the standard $1/r-$expansion in asymptotically flat spacetimes. 
This is one of the key differences between the study of linearized gravitational perturbations on de Sitter spacetimes and spacetimes for which $\Lambda=0$.
We assume that the physical size of the system is uniformly bounded by the cosmological radius $R_{\rm c}:=\sqrt{\frac{3}{\Lambda}}$, and that the physical velocity of the source is negligible, $v \ll 1$. 
Given these approximations, the gauge-invariant formula for power $P$ radiated by the system through any 2-sphere cross-section of $\scri^+$ orthogonal to the orbits of the time translation vector field $T^a$ is
\begin{equation}
\label{eq:power}
P \hat{=} \frac{G}{8 \pi} \int d^2 S \left[ \mathcal{R}^{ab} (\vec{x}) \; \mathcal{R}^{TT}_{ab} (\vec{x}) \right], 
\end{equation} 
where the radiation field $\mathcal{R}_{ab}$ is given by
\begin{equation}
\label{eq:radiationfield}
\mathcal{R}_{ab}( \vec{x}) \hat{=} \left[\dddot{Q}_{ab}^{(\rho)} + \sqrt{3 \Lambda} \ddot{Q}_{ab}^{(\rho)} + \frac{2 \Lambda}{3} \dot{Q}^{(\rho)}_{ab} 
+  \sqrt{\frac{\Lambda}{3}} \ddot{Q}_{ab}^{(p)} + \Lambda  \dot{Q}_{ab}^{(p)} + 2 \left(\frac{\Lambda}{3}\right)^{3/2} Q^{(p)}_{ab} \right] \,(t_{\rm ret}) ,
\end{equation}
which on $\scri^+$ is a function of $r$ only after taking the late time limit of the retarded time $t_{\rm ret}= - \sqrt{\frac{3}{\Lambda}} \ln (\sqrt{\frac{\Lambda}{3}} (-\eta+r))$ . 
The $\hat{=}$ denotes equality on $\scri^+$ and overdots represent Lie derivatives along $T^a$.
The label $TT$ refers to the transverse-traceless part of the radiation field, so that the divergence as well as the spatial trace of $\mathcal{R}_{ab}^{TT}$ vanish. 
Details on how to calculate the transverse-traceless part of any spatial, symmetric rank-2 tensor 
are discussed in the Appendix. 
It requires solving a set of spherical Poisson equations and \textit{cannot} be replaced by the algebraic projection operator that projects spatial tensors orthogonal to the radial direction, as is typically 
done for perturbations on Minkowski spacetime. 
This is because the algebraic projection operator is local in space, whereas the transverse-traceless part of a spatial tensor is global in space. 
Consequently, these two notions are completely unrelated.
Surprisingly, for asymptotically flat spacetimes, the radiative degrees of freedom on $\scri^+$ can in fact be extracted using either notion (for  a detailed comparison between the two notions in the context of 
asymptotically flat spacetimes, see \cite{ab1,ab2}). This is not true for de Sitter spacetimes. 
Although --- as we shall see in Sec.~\ref{sec:powerresult} --- for the specific example of a binary system in a circular orbit, the projected radiation field and the transverse-traceless part of the radiation field are the same in the high-frequency approximation, despite being drastically different before taking the high-frequency approximation. 
Whether this generalizes to other scenarios is not clear at the moment. 

Some of the salient properties of the above expression for power radiated, which  will also be important in the next section, are:

(1) Since the label $TT$ is only on one of the radiation fields and not on both, 
the expression is not manifestly positive. 
Nonetheless, in \cite{abk3}, it was proven that 
the power radiated is indeed positive. \\

\begin{figure}
\begin{center}
\includegraphics[scale=0.25]{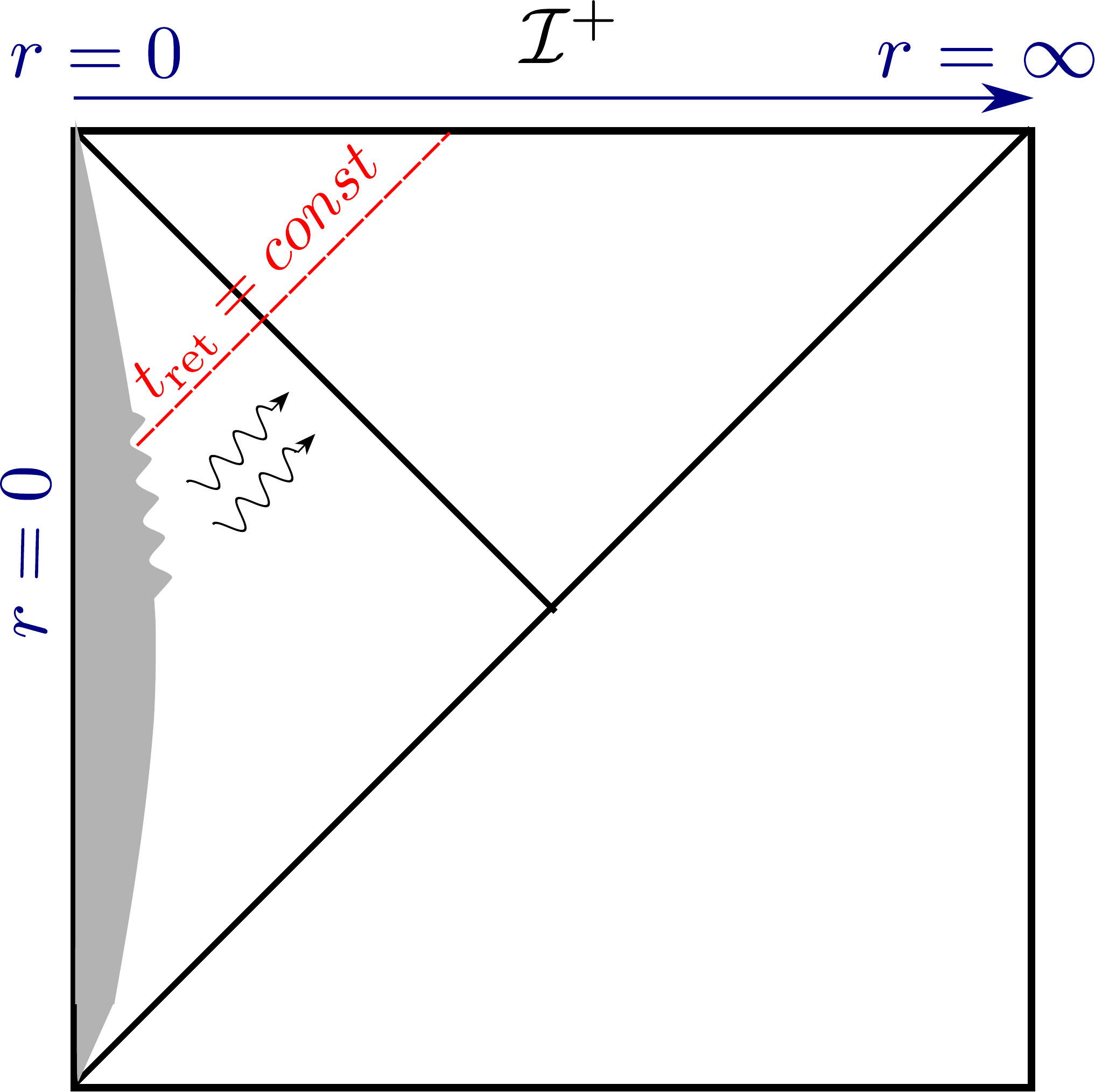}
\end{center}
\caption{\label{fig:conformaldiagram} This conformal diagram shows a source that is static for most times and dynamical between $t_1 <t <t_2$ during which the source emits gravitational radiation. 
This radiation is registered on $\scri^+$. 
The coordinate $u$ is the retarded time coordinate and represents the null direction along which gravitational perturbations travel. 
The power through any 2-sphere cross-section orthogonal to the orbits of the Killing vector field $T^a$ on $\scri^+$ is completely determined by the dynamics of the source at the corresponding retarded time.}
\end{figure}

(2) The expression for radiated power only involves radiation fields evaluated at retarded time. 
This is critical since this means that
the time scales in the radiation field are set by the wavelength evaluated at the source, and \emph{not} by the physical wavelength in the asymptotic region. 
Consequently, despite the fact that the physical wavelength of the gravitational field perturbations grow exponentially as the wave propagates and even exceed the cosmological radius $R_{\rm c}$ in the asymptotic region near $\scri^+$, the power in de~Sitter spacetimes is not `diluted' on $\scri^+$. 
This is illustrated in Figure~\ref{fig:conformaldiagram}.\\

(3) An interesting property of gravitational waves generated on de~Sitter spacetime is that the power carries information about the energy density \emph{and} the pressure of the source (if no assumption is made about the relative size of $\sqrt{\Lambda}$ and time derivatives of the quadrupole moments). 
This is different from the case when $\Lambda=0$, where the power to lowest post-Newtonian order is only proportional to the third time derivative of the mass quadrupole moment squared and pressure terms appear at higher order. \\

(4) Let us reiterate that the only approximations made in the derivation of \eref{eq:power} that describes the power radiated by a linearized source on a de Sitter background are: (i) the physical size of the source is smaller than the cosmological horizon, (ii) its velocity is small compared to the speed of light and (iii) the source is only dynamically active for a finite time period. (Although the later condition can most likely be relaxed and was only used to ensure finiteness of intermediate steps in the derivation.)
Thus, the radiative power presented in \eref{eq:power} does not rely on the high-frequency approximation (also known as the geometric optics or eikonal approximation). Nor is any temporal or spatial averaging performed to calculate the power radiated, as is explicit from the expression for power.
This is in contrast to standard expressions of power radiated on a flat background.

\section{Binary system in a circular orbit}
In this section, we first describe in detail the assumptions made to model the dynamics of the binary system in a circular orbit on a de Sitter background. This description applies to any FLRW background and is commonly used for these spacetimes. Here we review the treatment in standard references (that assume $\Lambda=0$) \cite{thorne, Maggiore:1900zz} and endeavor to be as explicit as possible with our assumptions\footnote{Thanks to Eric Poisson for clarifications in a private communication.}.
After this summary we show the result for radiated power of this system and comment on its properties. 
The details of the calculation of the transverse-traceless part of the quadrupole moments, needed for the calculation of power, can be found in the Appendix.

\subsection{Physical set-up}
\label{sec:physicalsetup}
To dynamically model the world lines of the two bodies with mass $m_1$ and $m_2$ making up the binary system, we make similar assumptions as in the standard treatment of sources in expanding spacetimes with $\Lambda=0$ \cite{thorne, Maggiore:1900zz}.
These assumptions are:
\begin{enumerate}
\item The characteristic (proper) time scale of the system $t_c$ and the expansion rate of the background are assumed to be such that the expansion of the Universe can be neglected during the orbital cycles of interest:
\begin{equation}\label{eq:HighFreq}
\frac{\dot{a}}{a} t_c \ll 1 \qquad \Longleftrightarrow \qquad \sqrt{\Lambda} t_c \ll 1.
\end{equation}
This is the familiar high-frequency (short-wave) approximation\cite{Isaacson:1968}\footnote{Technically, the high-frequency approximation is formulated in terms of conformal time $\eta$: $\mathcal{H} \eta_c \ll 1$ where $\mathcal{H}= \frac{1}{a} \frac{\del a}{\del \eta}$ and $\eta_c$ is the characteristic conformal time. 
Given $t_c$, however, we can relate this to the characteristic conformal time by $\eta_c \simeq \frac{t_c}{a_e}$ (where $a_e$ is the scale factor at the time of emission). 
Since $ \mathcal{H} \eta_c \simeq H \, t_c \,a/a_e  $, $\mathcal{H}(\eta) \eta_c \ll 1$ is the same as $H t_c \ll 1$ when we evaluate expressions at the time of emission. 
For a de Sitter background, the Hubble parameter is $H= \sqrt{\frac{\Lambda}{3}}$ so that the high-frequency approximation is equivalent to $\sqrt{\Lambda} t_c \ll 1$.}. 
\item The relative physical separation between the two bodies $R_* (=a r_* = a \left|\vec{r}_1 -\vec{r}_2\right|)$ is such that the bodies are far apart compared to the Schwarzschild radius of either body:
\begin{equation}
\frac{G m}{R_*} \ll 1 \qquad {\rm and} \qquad \frac{G \mu}{R_*} \ll 1 \, ,
\end{equation}
where the total mass of this system is $m=m_1+m_2$ and the reduced mass $\mu= \frac{m_1 m_2}{m}$. 
\item Each body moves slowly, that is, in the center of mass frame $v \ll 1$.
\item The pressure of each body is negligible compared to the energy density.
\item Each body is approximately spherically symmetric. 
\item The trajectory of the binary is well approximated by a circular orbit\footnote{The motivation for such a binary is strong since gravitational waves effectively act to make eccentric systems circular over time \cite{Peters:1964}, and sources seen later in their evolution are expected to have small eccentricity \cite{Hinder:2007qu}.}.
\end{enumerate}

The first assumption ensures that the time evolution of the scale factor can be neglected during the few cycles the system is studied. 
Therefore, the time behavior of the physical separation $R_*$ a few orbits before and after the time of emission is governed by the time behavior of the conformal separation $r$ multiplied by the scale factor at the time of emission $a_e$. 
Similarly, the physical angular velocity $\Omega$ is described by the conformal angular velocity $\omega$ multiplied by $a_e^{-1}$:
\begin{eqnarray*}
R_*(t) &=& a_e r_* + \Or (\sqrt{\Lambda} t_c), \\
\Omega(t) &=& a_e^{-1} \omega + \Or (\sqrt{\Lambda} t_c).
\end{eqnarray*}
From the first two assumptions, it also follows that by a simple constant rescaling of the coordinates, the spacetime metric is well approximated by a Minkowski metric  during the few cycles the system is studied:\footnote{As a side remark: since the scale factor is approximately constant, the Hubble radius is infinite to zeroth approximation during this period.}
\begin{equation}
ds^2 = - d\tilde{\eta}^2 + d\tilde{r}^2 + \tilde{r}^2 \left(d\theta^2 + \sin^2 \theta \; d \varphi^2 \right) + \Or (\sqrt{\Lambda} t_c)
\label{eq:rescaledmink}
\end{equation}
where $\tilde{\eta} := a_e \eta$ and $\tilde{r}:= a_e r$.

Given these rescaled coordinates, one can now interpret approximation 2-6 as the Newtonian approximation and use Newtonian dynamics to describe the motion of the binary in the rescaled coordinates.
Approximation 1 is critical for this interpretation.
An example of a binary system satisfying all of the above approximations is a binary consisting of two weakly self-gravitating bodies such as main-sequence stars. 
In addition, other examples are binaries of two compact bodies such as  neutron stars or black holes.
The internal gravity of each body can be arbitrarily strong but as long as their mutual gravity is weak, the Newtonian description applies.

\begin{figure}
\begin{center}
\includegraphics[scale=0.3]{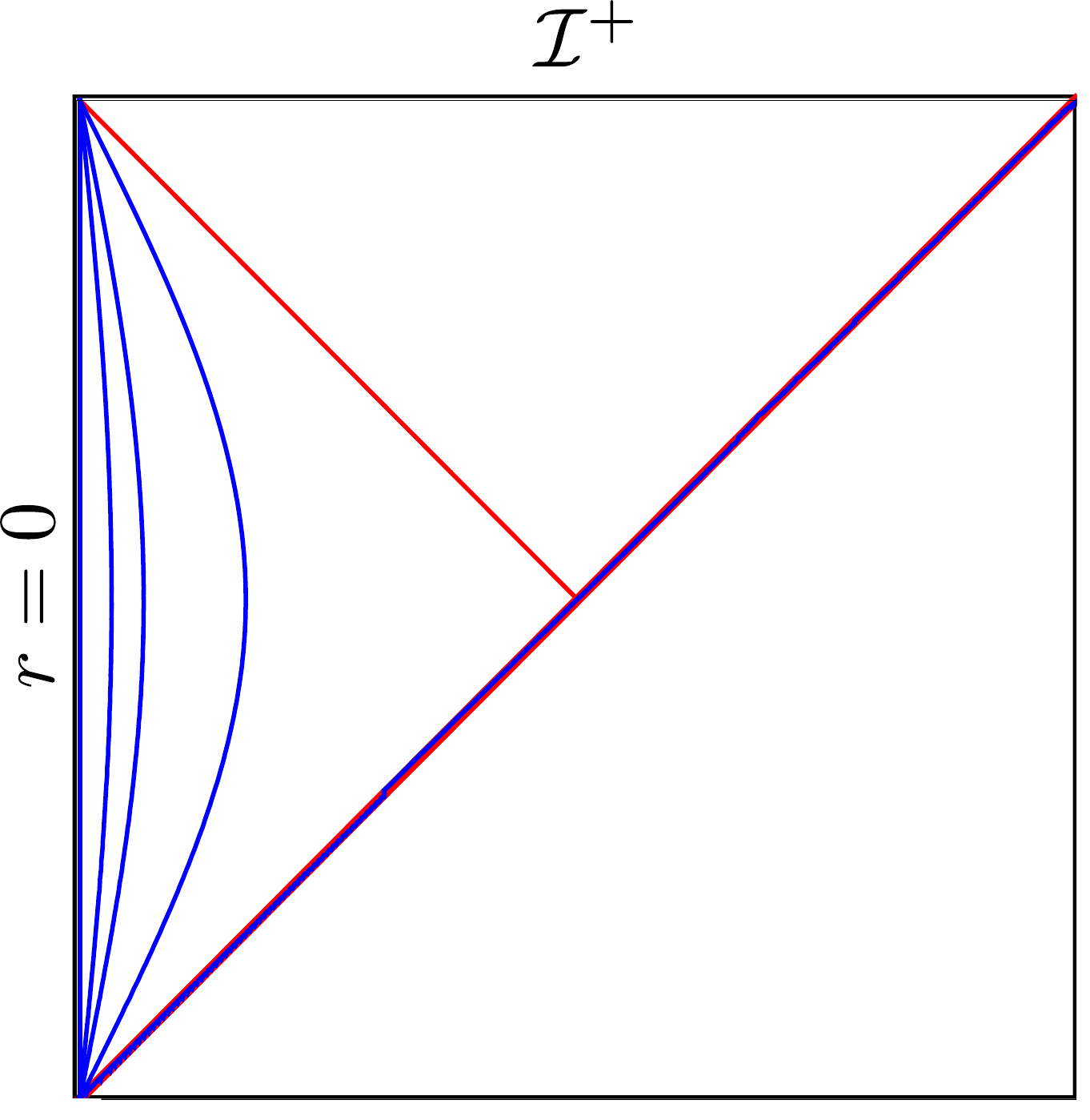}
\end{center}
\caption{\label{fig:conformalphysical} This conformal diagram illustrates the physical set-up for a binary 
system in a circular orbit.  The straight 
vertical (blue) line on the left indicates the origin $r=0$ around which the reduced mass 
of the binary system revolves. The curved (blue) lines are -- from left to right 
-- the trajectories of the reduced mass at $R_*=\frac{1}{10} R_{\rm c}, 
\frac{1}{5} R_{\rm c}$ and $ \frac{1}{2} R_{\rm c}$. The values for these trajectories are chosen for illustrative purposes; realistic systems will have a much smaller $R_*$. The shorter diagonal (red) line denotes the cosmological horizon of the source. (Note that in this figure $\frac{dR_*}{dt}$ is assumed to be exactly zero; not just $\frac{dR_*}{dt} = \Or(\sqrt{\Lambda}\Omega^{-1})$.) }
\end{figure}

In Newtonian mechanics, the motion of the center of mass is uniform. 
Consequently, the description of the motion simplifies when the origin of the coordinate system is chosen such that it is attached to the center of mass; the position of each body can then be determined in terms of the relative separation between the bodies. 
This is what we will do here as well.
From conservation of angular momentum, it follows that the orbital motion proceeds within a fixed orbital plane. 
Choosing the orbital plane to coincide with the $x-y$ plane and introducing the orbital angle to be $\varphi$, the equations of motion are
\begin{align}
&\frac{d^2 \tilde{r}}{d \tilde{\eta}^2} = \left(\tilde{r} \left(\frac{d \varphi}{d \tilde{\eta}}\right)^2 - \frac{G m}{\tilde{r}^2}  \right) \left[1 + \Or(\sqrt{\Lambda}t_c) + \Or\Big(\frac{G m}{R_*}\Big) + \Or\Big(\frac{G \mu}{R_*}\Big) \right] \; ,\label{eq:newton1}\\
&\frac{d}{d\tilde{\eta}} \left( \tilde{r}^2 \frac{d \varphi}{d \tilde{\eta}} \right) = 2 G m \,\left( 2 - \frac{\mu}{m} \right)  \frac{d \tilde{r}}{d \tilde{\eta}} \frac{d \varphi}{d \tilde{\eta}}  \left[1+ \Or(\sqrt{\Lambda}t_c) + \Or\Big(\frac{G m}{R_*}\Big) + \Or\Big(\frac{G \mu}{R_*}\Big) \right]
\label{eq:newton2}
\end{align}
For a circular orbit $\frac{d \tilde{r}}{d \tilde{\eta}} =0$ and it follows from \eref{eq:newton2} that $\frac{d \varphi}{d \tilde{\eta}} \equiv a_e^{-1} \omega$ is constant. 
In other words, the physical angular velocity $\Omega$ is constant up to terms proportional to $\Or (\sqrt{\Lambda} t_c)$. 
From \eref{eq:newton1}, we obtain Kepler's law:\footnote{This is the same Kepler's law that is used for FLRW spacetimes. 
Often it is written in a slightly different format in which the above physical quantities defined at the source are related to quantities as measured by an observer at redshift $z$: $\omega^{\rm obs} = (1 + z)^{-1} a_e^{-1} \omega$, $m_z = (1 + z) m$ and $r_z = (1+z) a_e r$. In terms of the observed quantities --- given the same approximations --- Kepler's law reads:
\begin{align}
\omega_{\rm obs}^2 = \frac{G m_z}{r^3_z}.
\end{align}
}
\begin{align}
\left(\frac{d \varphi}{d \tilde{\eta}}\right)^2 = \frac{G m}{\tilde{r}^3} \left[1 + \Or(\sqrt{\Lambda} t_c) + \Or\Big(\frac{G m}{R_*}\Big) + \Or\Big(\frac{G \mu}{R_*}\Big) \right].
\end{align}
In terms of the physical separation vector $R_*$ and angular frequency $\Omega$, Kepler's law is:
\begin{align}
\Omega^2 = \frac{G m}{R_*^3} \left[1 + \Or(\sqrt{\Lambda} t_c) + \Or\Big(\frac{G m}{R_*}\Big) + \Or\Big(\frac{G \mu}{R_*}\Big) \right].
\label{eq:Keplerslaw}
\end{align}
Once truncated this version of Kepler's law is similar in form to the Minkowski version, with the only difference being constant factors of $a_e$.
The trajectory of the reduced mass as a function of proper time is now described by 
\begin{align*}
\vec{R}_*(t) &= \left(R_*(t) \, \cos \left(\Omega t + \frac{\pi}{2} \right) \hat{x} + R_*(t) \, \sin \left(\Omega t + \frac{\pi}{2} \right) \hat{y} \right)  \left[1 + \Or(\sqrt{\Lambda} t_c) + \Or\Big(\frac{G m}{R_*}\Big) + \Or\Big(\frac{G \mu}{R_*}\Big) \right] 
\end{align*}
where $\hat{x}$ ($\hat{y}$) is the unit vector in the $x-$direction ($y-$direction) with respect to the Euclidean 3-metric and $\frac{dR_*}{dt}= \Or (\sqrt{\Lambda} t_c)$ and $\frac{d\Omega}{dt} = \Or(\sqrt{\Lambda}t_c)$.
The trajectory of the reduced mass is illustrated in Figure~\ref{fig:conformalphysical} for various choices of $R_*$. 
The energy density of this system is 
\begin{eqnarray*}
\rho &=& \mu  \, \delta^{(3)}(\vec{R}-\vec{R}_*) \left[1 + \ord (\sqrt{\Lambda} t_c) + \Or\Big(\frac{G m}{R_*}\Big) + \Or\Big(\frac{G \mu}{R_*}\Big)\right] \\
&=& \frac{\mu}{a_e^3} \delta^{(3)}(\vec{r}-\vec{r}_*) \left[1 + \ord (\sqrt{\Lambda} t_c)+ \Or\Big(\frac{G m}{R_*}\Big) + \Or\Big(\frac{G \mu}{R_*}\Big) \right].
\end{eqnarray*}
This choice for the energy density ensures that the reduced mass $\mu$ remains constant.
Note that the stress-energy tensor $T_{ab}$ constructed from this energy density is only approximately conserved.
In other words, the divergence of the stress-energy tensor with respect to the derivative operator compatible with the de Sitter metric is $\ord (\sqrt{\Lambda} t_c)$: $\bar{\grad}^a T_{ab} = \ord (\sqrt{\Lambda} t_c)$.
With these choices, the (mass) quadrupole moment is given by
\begin{align}
Q_{ab}^{(\rho)}(t) & = \frac{\mu R_*^2(t)}{2} \left[ \left(1 - \cos 2 \Omega t \right) \grad_a x \grad_b x + \left(1 + \cos 2 \Omega t \right) \grad_a y \grad_b y \right.
\notag \\ 
&\qquad  
\left. - 2\sin 2 \Omega t \; \grad_{(a} x \grad_{b)} y \right] 
\left[1 + \Or(\sqrt{\Lambda} t_c) + \Or\Big(\frac{G m}{R_*}\Big) + \Or\Big(\frac{G \mu}{R_*}\Big) \right] \, .
\label{eq:quadphys}
\end{align}
Taking $\Lambda \to 0$ is equivalent to taking $a \to 1$ and the quadrupole moment reduces to the flat space result in this limit \cite{Peters:1963ux}.

\subsection{Power radiated}
\label{sec:powerresult}
Given this set-up, we can now proceed to calculate the radiated power of this system and comment on its properties. 
The details of the calculation of the transverse-traceless part of the quadrupole moments, needed for the calculation of power, can be found in the Appendix.

Once the transverse-traceless decomposition of the quadrupole moment $Q_{ab}^{(\rho)}(t_{\rm ret})$ in \eref{eq:quadphys} and its derivatives, are calculated (for explicit expressions, see eqs.(\ref{eq:QTTrr})-(\ref{eq:QTTphiphi})), we use the leading order part in the high-frequency approximation of eq.(\ref{eq:power}) to calculate the radiated power:
\begin{align}
P &\hat{=} \frac{G}{8 \pi} \int d^2 S \; \dddot{Q}^{\rm TT}_{ab} \dddot{Q}^{ab} \left[1 + \Or(\sqrt{\Lambda} \Omega^{-1}) \right]\label{eq:powerdShighfreq}\\
&\hat{=} \frac{32 G}{5} \mu^2 R_*^4(t_{\rm ret}) \Omega^6(t_{\rm ret}) \left[1 + \Or(\sqrt{\Lambda} \Omega^{-1}) + \Or\Big(\frac{G m}{R_*}\Big) + \Or\Big(\frac{G \mu}{R_*}\Big) \right] \; ,
\end{align}
where, as before, $\hat{=}$ denotes equality on $\scri$.
Since $R_*$ and $\Omega$ are constant in time given the approximations made, this equation can simply be written as:
\begin{align}
P &\hat{=} \frac{32 G}{5} \mu^2 R_*^4 \Omega^6 \left[1 + \Or(\sqrt{\Lambda} \Omega^{-1}) + \Or\Big(\frac{G m}{R_*}\Big) + \Or\Big(\frac{G \mu}{R_*}\Big) \right].
\label{eq:resultpower}
\end{align}
Let us first comment on three of the main features of this formula. Afterwards, we will contrast this result to the power radiated by a binary system on Minkowski spacetime.
First, the radiated power is manifestly gauge-invariant as it solely depends on physical quantities. In addition, the power is clearly positive. 
Positivity of the power radiated in de~Sitter spacetimes was proven for gravitational waves generated by physically realistic sources in \cite{abk3}. 
This is the first explicit illustration of that general proof.
Third, since there are only two independent scales in this system, the mass and  distance scale, the expression for power can be simplified further. 
Using Kepler's law, see \eref{eq:Keplerslaw}, the expression for radiated power on a de~Sitter background reduces to
\begin{align}
P &\hat{=} \frac{32}{5} \frac{1}{G} \left(G M_c \Omega \right)^{10/3}  \left[1 + \Or(\sqrt{\Lambda} \Omega^{-1}) + \Or\Big(\frac{G m}{R_*}\Big) + \Or\Big(\frac{G \mu}{R_*}\Big) \right] \; ,
\end{align}
where $M_c:= \mu^{3/5} m^{2/5}$ is the chirp mass.

\begin{figure}
\begin{center}
\includegraphics[scale=0.25]{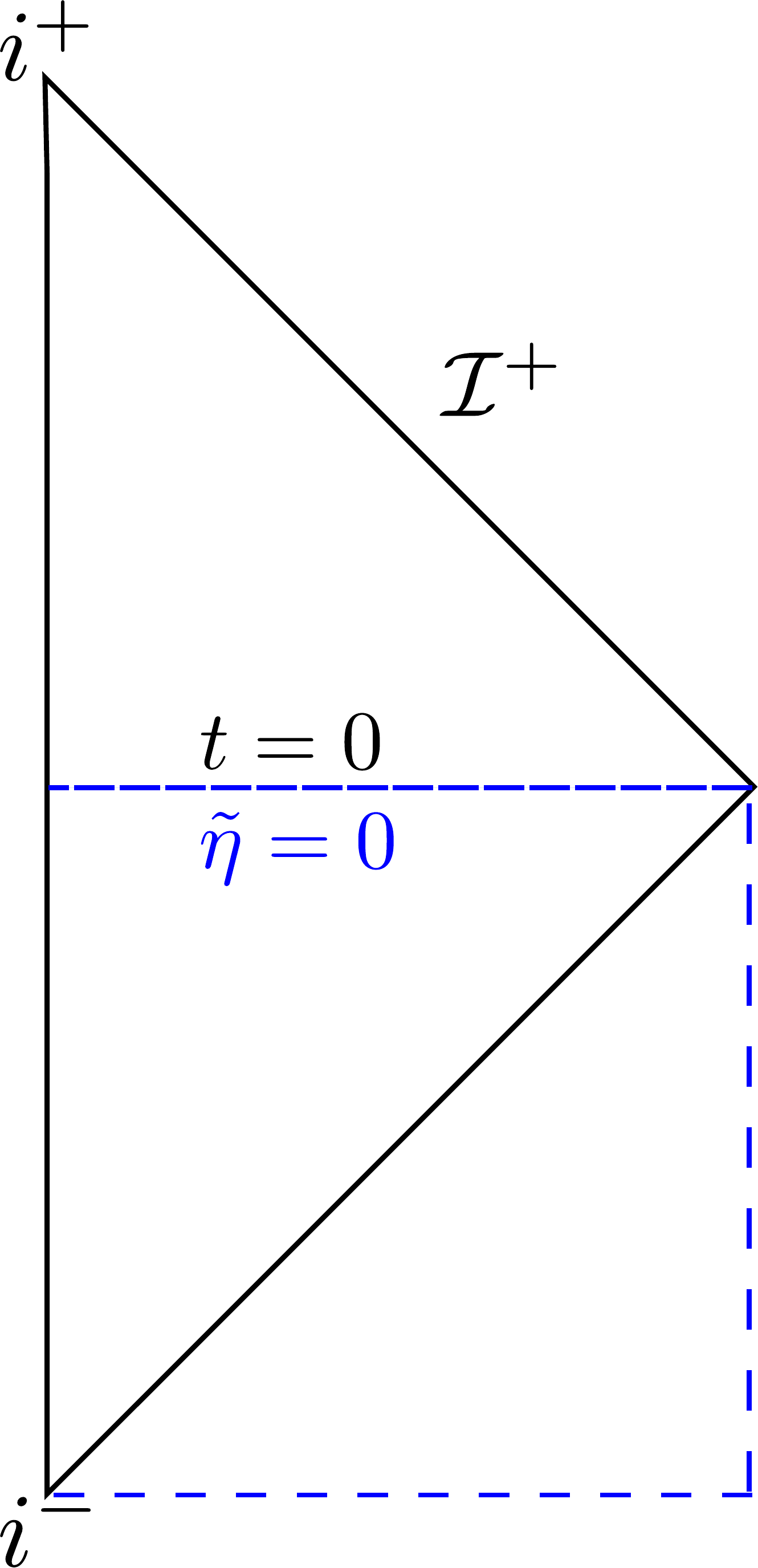}
\end{center}
\caption{\label{fig:embedding} This diagram shows the $\tilde{\eta}=0$ surface of the rescaled flat coordinates. From a Minkowskian perspective, this surface is simply the $t=0$ surface in the middle of Minkowski spacetime. From a de Sitter perspective, this surface is its future null infinity $\scrip$.}
\end{figure}

The result in Minkowski spacetime is identical to eq.~\eqref{eq:resultpower} up to constant factors of $a_e$. 
This is surprising for several reasons.
At the conceptual level, it seems that de Sitter calculation is essentially carried out on the rescaled Minkowski spacetime in \eref{eq:rescaledmink}. 
However, \emph{the formula for power used is defined on the de Sitter $\scri$ and with respect to the de Sitter time translation}.
In the rescaled flat coordinates, $\scri$ of de Sitter spacetime is at the $\tilde{\eta}=0$ time slice, which from the Minkowskian perspective is a space-like slice in the middle of spacetime.
This is illustrated in Figure~\ref{fig:embedding}.
In addition, the de Sitter time translation is a space-like conformal Killing vector field on the $\tilde{\eta}=0$ surface in the rescaled flat coordinates.
How then is it possible that this power is equivalent to the power radiated across $\scri$ of Minkowski spacetime that is defined with respect to the time translation Killing field of Minkowski space, which is time-like everywhere?
Furthermore, the calculational techniques used to derive the power radiated on a de Sitter background are drastically different from those on a Minkowski background. The power radiated in de~Sitter spacetime relies on a late time expansion to reach $\scri^+$, whereas the power radiated in Minkowski spacetime uses a $1/r$ expansion.
In addition, the calculation of the transverse-traceless part on a de Sitter background cannot be done using the algebraic projection operator, instead one is required to solve a set of differential equations to extract the transverse-traceless part. 
An explicit calculation shows that, for the binary system on a de Sitter background discussed here, $Q_{ab}^{TT}$ and $Q_{ab}^{tt}$ are indeed different (and this difference does \emph{not} vanish in the high-frequency limit):
\begin{align}
\label{eq:diffTTvstt}
Q_{ab}^{TT}-Q_{ab}^{tt} &\hat{=} 
\frac{\mu R_*^2}{2} \left(
- \frac{1}{6} \left(1+3 \cos 2\theta \right) \; \grad_a r \grad_b r + r \sin 2 \theta \; \grad_{(a} r \grad_{b)} \theta  \right. \notag \\
& \quad \left. + \frac{r^2}{12}  \left(1+3 \cos 2\theta \right) \grad_a \theta \grad_b \theta + \frac{r^2}{12} \sin^2 \theta  \left(1+3 \cos 2\theta \right) \grad_a \varphi \grad_b \varphi \right) \left[1 + \Or (\sqrt{\Lambda} \Omega^{-1}) \right]
\end{align}
where $Q_{ab}^{tt}$ refers to the algebraically projected quadrupole moment
\begin{align}
Q_{ab}^{tt} := \left( P_a^{\, c} P_b^{\,d} - \frac{1}{2} P_{ab} P^{cd} \right) Q_{cd}
\end{align}
with $P_a^{\;b} = q_a^{\, b} - \grad_a r \; \partial_r^b$ and satisfying $P_a^{\,b} r_b=0$,  $P_a^{\,a}=2$, and $P_a^{\,c} P_c^{\,b} = P_a^{\,b}$.
Moreover, the power on a flat background typically makes use of some type of spatial or temporal averaging, such as Brill-Hartle averaging. 
No such averaging appears in the formula for power radiated on a de Sitter background.

This puzzle is resolved by carefully considering the power radiated across the $t=0$ surface in Minkowski spacetime.
Assuming that the source was only dynamic for a finite time interval before $t=0$, the formula for power radiated at the spatial surface given by $t=0$ is
\begin{equation}
P_{\rm Mink}(r) \hat{=} \frac{G}{8 \pi} \int_{t=0} d^2 S \; r^2 \left[ \left(\frac{\dddot{Q}_{ab}(t_{\rm ret})}{r}\right)^{TT} \frac{\dddot{Q}_{ab}(t_{\rm ret})}{r} + \ord \Big( \frac{1}{r^3} \Big) \right] \;, 
\label{eq:powerMinkt0}
\end{equation}
where the overdots  refer to the time translation vector field of Minkowski spacetime.
Note that since this formula is not evaluated at $\scri$ of Minkowski spacetime, the radiative degrees of freedom in $\frac{\dddot{Q}_{ab}(t_{\rm ret})}{r}$ cannot be obtained by simply projecting it onto the two-sphere.
This procedure is only valid on Minkowski $\scri$.
Therefore, the factor of $r^{-1}$ cannot be pulled out of the TT-operation and the $r$'s in this formula do not cancel at leading order.
Together with the fact that there are also higher order terms that cannot be neglected for finite $r$, this shows that the formula across the $t=0$ surface in Minkowski spacetime in eq.~\eqref{eq:powerMinkt0} is rather different from the de Sitter formula across the Minkowskian $\tilde{\eta}=0$ surface in eq.~\eqref{eq:powerdShighfreq}.
Hence, not only are the Killing vector fields used to define the power radiated across the Minkowskian $t=0$ versus $\tilde{\eta}=0$ surface 
different, also the expressions for power are. 
Only in the limit to $\scri$ does the power radiated in Minkowski spacetime become 
\begin{equation}
P_{\rm Mink} \hat{=} \frac{G}{8 \pi} \int_{\scri} d^2 S \; \dddot{Q}_{ab}^{\rm tt}(t_{\rm ret}) \dddot{Q}_{ab}(t_{\rm ret}) 
\label{eq:powerMinkscri}
\end{equation}
and does the power radiated in Minkowski spacetime agree with the power across $\tilde{\eta}=0$ of the rescaled flat spacetime using the de Sitter expression.
In summary, the power radiated by a binary satisfying approximation 1-6 across $\scri$ of de Sitter spacetime is equal to the power radiated by the same binary across $\scri$ of Minkowski spacetime.
\\

So far, we have focused on modeling a physically realistic binary system. 
As a result, the dynamics of the system could only reliably determined within the high-frequency limit and consequently the expression for power was only valid up to $\Or(\sqrt{\Lambda} \Omega^{-1})$.
Therefore, a natural question is: ``How does the power change if  the dynamics of the system is known to all orders in $\Or(\sqrt{\Lambda} \Omega^{-1})$?'' 
As a proof of principle, let us relax assumption 1 at the beginning of Sec.~\ref{sec:physicalsetup} and imagine that the source is in a circular orbit for \emph{all} time: 
\begin{equation}
\frac{d R_*}{d t}=0 \qquad {\rm and} \qquad \frac{d \Omega}{d t} =0.
\end{equation}
In other words, this assumes that the adiabatic approximation is valid on cosmological time scales. 
This is an externally specified, fine-tuned trajectory and not necessarily physical. 
(Although it does have the nice property that the bodies making up the binary system are not being pulled apart by the cosmological expansion and the system remains forever bounded by the short distance attraction.) 
If the associated energy density is the only non-zero component of the stress-energy tensor $T_{ab}$, the stress-energy tensor is not conserved on a de Sitter background and one cannot use eq.~\ref{eq:power} --- which relies on conservation of the stress-energy tensor --- beyond the high-frequency approximation.
However, there are many possible stress-energy tensors with non-zero space-time and space-space components such that $T_{ab}$ is conserved\footnote{Although it is unclear how realistic such stress-energy tensors are.}.
We consider a stress-energy tensor from this class with vanishing pressure.
Then using the results from the Appendix for the transverse-traceless part of the radiation field, the resulting power is:
\begin{align}
P &\hat{=} \frac{32 G}{5} \mu^2 R_*^4  \Omega^6  \left(1 + \frac{5}{12} \frac{\Lambda}{\Omega^2} + \frac{1}{36} \frac{\Lambda^2}{\Omega^4} \right) .
\label{eq:powerbeyondhighfreq}
\end{align}
This result highlights the fact that the power radiated can get corrections in a de Sitter spacetime. 
These corrections are linear and quadratic in $\Lambda$ with coefficients less than one. 
That these coefficients are small is a priori not obvious, especially given that the $\Lambda \to 0$ limit can be discontinuous as evidenced by an infinite set of solutions on de Sitter spacetime that carry negative energy while no such solutions exist on Minkowski spacetime \cite{abk3}.
The above expression for power is exact and not a truncation of a series in $\Lambda$. 
Taking the high-frequency limit of this result, recovers the result in \eref{eq:resultpower}. 
Quantitatively, the $\Lambda$-dependent corrections are extremely small for most physically realistic binary systems and unlikely to play a role in observations that measure power radiated directly. 
However, these corrections due to $\Lambda$  may result in an observable frequency shift for sources observed by future space-based missions. 
A more realistic system is needed to calculate the size of this effect.
\\

\textit{Remark.} Just as on a Minkowski background, for any isolated system, linear momentum radiated in the form of gravitational waves vanishes on a de Sitter background given the approximations made. 
For the specific case studied here, a binary system restricted to the $x-y$ orbital plane, the flux of angular momentum in the $x$- and $y$-direction also vanishes on Minkowski and de Sitter backgrounds. 
The instantaneous flux of angular momentum in the $z$-direction does not vanish and on de Sitter $\scri^+$ is given by:\footnote{Here, eq. (4.33) in \cite{abk3} was used.}
\begin{align}
\Lie_T J_z \hat{=}  \frac{32 G}{5} \mu^2 R_*^4 \Omega^5 
\left[1 + \Or(\sqrt{\Lambda} \Omega^{-1}) + \Or\Big(\frac{G m}{R_*}\Big) + \Or\Big(\frac{G \mu}{R_*}\Big) \right].
\end{align}

\section{Discussion}
\label{sec:discussion}
Previous results for linearized perturbations on a de Sitter background have shown certain effects compared to a flat spacetime background. 
Specifically, using different methods, several groups found that the gravitational memory effect is enhanced by redshift factors \cite{Chu:2015yua,Chu:2016qxp,BGY,TW}. This enhancement persists even after taking the high-frequency approximation. 
Thus, one might anticipate a similar enhancement for power radiated in the form of gravitational waves by a source on a de~Sitter background. 
This expectation is further supported by the striking differences between the calculational techniques used near $\scri^+$ of de~Sitter spacetimes and those used near $\scrip$ of Minkowski spacetimes: late time versus large $r$ expansions, extracting the radiative degrees of freedom by solving a set of differential equations versus using an algebraic projection operator and in the calculation of power, no averaging versus spatial or temporal averaging\footnote{Albeit, these differences do not indicate whether the power would be enhanced or diminished.}.
These expectations are not borne out. 
The power radiated by a binary system on de Sitter spacetime in terms of the reduced mass and angular velocity shows no enhancement (nor decrease) as compared to the result on a Minkowski background for systems we considered.
In other words, the standard expression for power radiated in Minkowski spacetimes also applies to de Sitter spacetime in the high-frequency approximation.
The high-frequency limit is critical for this equivalence. 

This result highlights that in order to probe the cosmological constant by measuring the power, one needs to go beyond the high-frequency approximation. 
Since the general expression for power radiated in de Sitter spacetimes does not invoke the high-frequency approximation, in principle, there are no obstacles to perform such a calculation.
In the current set-up, this regime could not be probed as the dynamics of the binary system could only be determined up to $\Or(\sqrt{\Lambda} \Omega^{-1})$.
Thus, if one could reliably determine the source dynamics beyond the high-frequency approximation, one could calculate the corrections due to the background curvature on the power. 
This would allow one to observe $\Lambda$ through the power emitted by gravitational waves. 

\begin{figure}
\begin{center}
\includegraphics[scale=0.25]{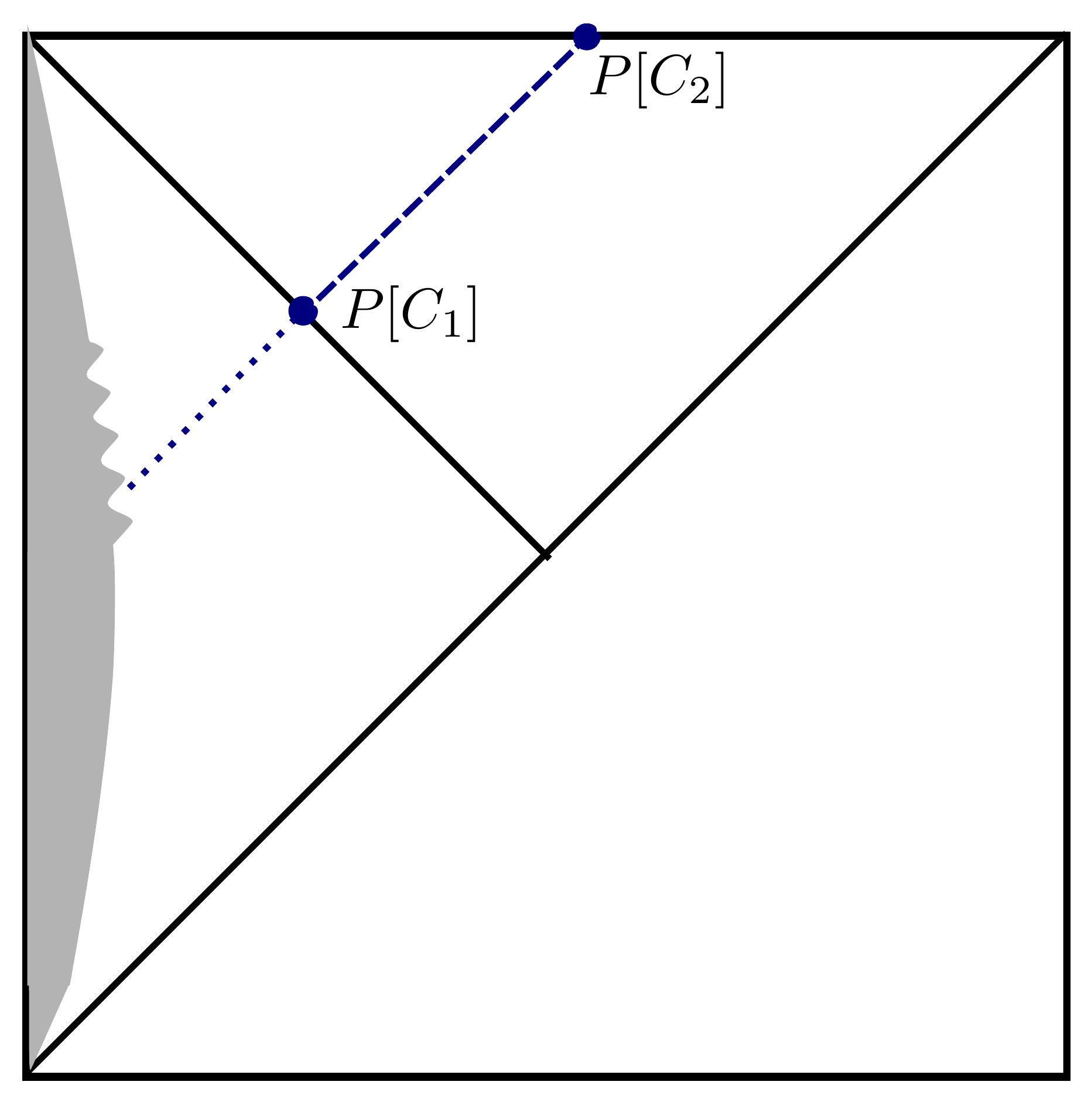}
\end{center}
\caption{\label{fig:extension} There are strong indications, \cite{Date:2016uzr}, that the power radiated across a 2-sphere cross-section on $\scri^+$ is the same as the power radiated across any 2-sphere along the light-ray emitted from $\scri^+$ to the source as long as one remains `far enough away from the source' and within the high-frequency approximation. 
Therefore, the power emitted across a 2-sphere cross-section on $\scri^+$, $P\left[C_2\right]$, is likely the same as the power emitted on the cosmological horizon, $P\left[C_1\right]$.}
\end{figure}

The power emitted by the binary system was calculated on $\scri^+$ of de Sitter spacetime. 
A natural question is: how does this result relate to what an observer at a finite time in de Sitter spacetime may detect? 
Since the dynamical part of the gravitational wave on a de Sitter background propagates sharply (its tail term is an `instantaneous' tail, see appendix A of \cite{abk3}), it is likely that the power radiated through a 2-sphere cross-section on $\scri^+$ is the same as it is through any two-sphere connected to the two-sphere on $\scri^+$ by a null ray \cite{Date:2016uzr}. 
In particular, the power radiated across $\scri^+$ is the same as it is through the cosmological horizon. 
This is illustrated in Figure~\ref{fig:extension}.
Thus, even though the power presented here was calculated on $\scri^+$, there are indications that this power is the same on the cosmological horizon, where an observer may detect such radiation.

\section{Acknowledgements}{
We are grateful to K.G. Arun, Abhay Ashtekar, Matt Benacquista, William East, Brajesh Gupt, Aruna Kesavan, Shane Larson, Luis Lehner, Eric Poisson, Bangalore Sathyaprakash, Charles Torre, James Wheeler and 
the strong gravity group at Perimeter Institute for fruitful discussions.
BB acknowledges support from NSF grant PHY-1505411, the Eberly research funds of Penn State and a Mebus Fellowship. JH is supported by the NSF NANOGrav Physics Frontiers Center, PHY-1430284. 
BB also would like to thank the Perimeter Institute for their hospitality, 
where parts of this project were completed.
}

\appendix
\section{Transverse-traceless decomposition}
\label{app:TT} 
In this Appendix, we first briefly recall the decomposition of a spatial tensor into its irreducible parts and outline a generic prescription to extract the transverse-traceless part of this decomposition. 
Next, we apply this method to calculate the transverse-traceless part of the quadrupole moment in \eref{eq:quadphys} and comment on the calculation of the transverse-traceless part of the radiation field $\mathcal{R}_{ab}$. For this calculation, the algebraic computing software Maple was used extensively, both for tensor analysis, using the package DifferentialGeometry \cite{Anderson:2011mb}, and solving these differential equations. \\

Any spatial symmetric rank-2 tensor $S_{ab}$ can be decomposed into 
its irreducible parts in the following form \cite{Cook2000,PoissonWill}:
\begin{align}\label{eq:Decomp}
S_{ab} = \frac{1}{3} q_{ab} S + \left(D_a D_b - 
\frac{1}{3} q_{ab} D^2 \right)B + 2 D_{(a} B_{b)}^T +
S_{ab}^{TT} 
\end{align}
where $q_{ab}$ is the spatially flat metric and $D_a$ is the covariant derivative compatible with $q_{ab}$. 
$S$ is the spatial trace of $S_{ab}$ ($S:=q^{ab}S_{ab}$), $B_a^T$ is a transverse vector so that $D^a B_a^T=0$, and $S_{ab}^{TT}$ is a transverse and traceless tensor\footnote{York, \cite{York:1973ia}, uses a similar decomposition using only the vector $W_a$, related to our components via $W_a=B_a^T+\frac{1}{2}D_aB$.}. 
Normally, in order to extract the radiative degrees of freedom of the gravitational field that are encoded in the transverse-traceless part of the spatial components, one often uses an algebraic projection operator in gravitational wave theory. 
However, as the above decomposition indicates, this generically will not extract the transverse-traceless part $S_{ab}^{TT}$. Instead one needs to solve a set of differential equations. 
The ``recipe'' to obtain the TT part of a tensor $S_{ab}$ is to first take one and two covariant derivatives of \eref{eq:Decomp}. This results in two Poisson equations and one vector-Poisson equation \cite{Cook2000}. First, one  solves for $A$ in $D^2 A = \frac{3}{2} D^a D^b S_{ab} - \frac{1}{2} D^2 S$. Second, solve for $B$ in $D^2 B = A$. Third\footnote{Note that the second and third steps can be interchanged.}, solve for $B^T_a$ satisfying
		\begin{align}
		D^2 B^T_a &= D^b S_{ab} - \frac{1}{3} D_a S - \frac{2}{3} D^2 D_a B \; , 
		\label{eq:d2b}\\
		D^a B_a^T &=0. \label{eq:transverse}
		\end{align}
Knowing $S$, $B$ and $B_a^T$, one obtains $S_{ab}^{TT}$ by subtracting these components from $S_{ab}$ using \eref{eq:Decomp}.

Now we apply this procedure to calculate the transverse-traceless part of the quadrupole moment in \eref{eq:quadphys}. In each step, we only keep the particular solutions to the Poisson equations as we want the solution due to the source of the gravitational waves. 
In the first and second steps normal techniques for solving the Poisson equation can be used. 
Since the source is quadrupolar finding solutions is straightforward. 
In spherical coordinates the radial component of the vector-Poisson equation in \eref{eq:d2b} is simplified by using the transverse condition to substitute out all but the radial components of the vector-Laplacian operator, giving an almost-standard scalar Helmholtz equation. 
The $\theta$-component of the vector-Poisson equation is the most challenging, since the source term for the component that appears on the right hand side of \eref{eq:d2b} is not purely quadrupolar.
Fortunately, the angular dependence of the solution follows from the source term, so using an ansatz of the form
\begin{align}
B_\theta&= \mu R_*^2  \sin 2\theta  \bigg[f_1(r)\cos \left(2 \varphi - 2 \frac{\Omega}{H}\ln(Hr)\right) + f_2(r) \sin \left(2 \varphi - 2  \frac{\Omega}{H}\ln(Hr)\right) \bigg]
\end{align}
allows one to solve for the two functions of the radial coordinate, $f_1(r)$ and $f_2(r)$, and obtain a solution. 
Here the index $\theta$ refers to the orthonormal component of $B_a^T$ in the direction of $\theta$. (Similary, we shall use $r$ and $\varphi$ indices to denote the orthonormal components proportional to $\grad_a r$ and $\grad_a \varphi$, respectively.) 
Once these two components of $B_a^T$ are known, the transverse condition can be integrated to find the third component, i.e. $B_\varphi^T$.
This results in the following transverse-traceless components of the quadrupole moment $Q_{ab}$ evaluated at retarded time on de Sitter $\scri^+$ (again, written in an orthonormal frame):
\begin{align}
Q^{TT}_{rr} &\hat{=}-\frac{3 \mu  R_*^2 H^2 \sin ^2\theta }{\left(9
   H^2+4 \Omega ^2\right) \left(25 H^2+4 \Omega ^2\right)}\Big[\left(15 H^2-4 \Omega ^2\right) \cos \left(2 \Omega  t_{\text{ret}}+ 2 \varphi \right)
 +16 H \Omega  \sin \left(2 \Omega  t_{\text{ret}}+ 2 \varphi \right)
    \nonumber
    \\   
   & \quad +\left(9 H^2+4 \Omega ^2\right) \left(25 H^2+4 \Omega
   ^2\right)		
   	+\cos (2 \theta ) \left(675 H^4+408 H^2 \Omega ^2+48 \Omega ^4\right)\Big] \label{eq:QTTrr}
\\
Q^{TT}_{r\theta} &\hat{=}\frac{\mu  R_*^2 \sin (2 \theta )}{4 \left(25 H^2+4 \Omega ^2\right)} \Big[-4 H \Omega  \sin \left(2 \Omega  t_{\text{ret}}+ 2 \varphi \right) - 10 H^2 \cos \left(2 \Omega  t_{\text{ret}}+ 2 \varphi \right) +25 H^2+4 \Omega ^2\Big]
\\
Q^{TT}_{r\varphi} &\hat{=}\frac{\mu  R_*^2 H\sin \theta}{25 H^2+4 \Omega ^2} \left[5 H \sin \left(2 \Omega  t_{\text{ret}}+ 2 \varphi \right)-2 \Omega  \cos \left(2 \Omega  t_{\text{ret}}+ 2 \varphi \right) \right]
\\
Q^{TT}_{\theta \theta} &\hat{=}\frac{\mu R_*^2 }{12 \left(9 H^2+4 \Omega ^2\right) \left(25 H^2+4 \Omega ^2\right)}
\Big[ 
-16 \Omega ^4 - 136 H^2 \Omega ^2 -225 H^4 
 +\cos (2 \theta ) \left(675 H^4+408 H^2 \Omega ^2+48 \Omega ^4\right) 
\notag 
\\ & \quad -6 \left(45 H^4+87 H^2 \Omega ^2 
+\cos (2 \theta ) \left(45 H^4+21 H^2 \Omega ^2+4 \Omega ^4\right)+12 \Omega ^4\right) 
\cos \left(2 \Omega  t_{\text{ret}}+ 2 \varphi \right)  \notag \\
& \quad +6 H \Omega  \left(\cos (2 \theta ) \left(4 \Omega ^2-3 H^2\right)+87 H^2+12 \Omega ^2\right) \sin \left(2 \Omega  t_{\text{ret}}+ 2 \varphi \right)  
\Big]
\\
   Q^{TT}_{\theta \varphi} &\hat{=}\frac{\mu  R_*^2 \cos (\theta )}{225H^4+136 H^2 \Omega ^2+16 \Omega ^4} \Big[\left(42 H^3 \Omega +8 H \Omega ^3\right) \cos \left(2 \Omega  t_{\text{ret}}+ 2 \varphi \right)
   \nonumber\\
   			&
   			+\left(45 H^4+54 H^2 \Omega ^2+8 \Omega
   ^4\right) \sin \left(2 \Omega  t_{\text{ret}}+ 2 \varphi \right)\Big]
\\
   Q^{TT}_{\varphi \varphi} &\hat{=}\frac{\mu  R_*^2}{12 \left(9 H^2+4 \Omega ^2\right) \left(25 H^2+4 \Omega ^2\right)} \Bigg[
   450 H^4+272 H^2 \Omega ^2+32 \Omega ^4 \nonumber \\
   			& \quad  -6  H \Omega \left(\cos (2
   \theta ) \left(45 H^2+4 \Omega ^2\right)+39 H^2+12 \Omega ^2\right) \sin \left(2 \Omega  t_{\text{ret}}+ 2 \varphi \right)  \nonumber\\
   			& \quad 			6 \left(90 H^4+\Omega ^2 \cos (2 \theta ) \left(33 H^2+4 \Omega ^2\right)+75 H^2 \Omega ^2+12 \Omega ^4\right) \cos \left(2 \Omega  t_{\text{ret}}+ 2 \varphi \right)
   			\Bigg] \label{eq:QTTphiphi} \;.
\end{align}
Let us comment on a few properties of the above expressions. 
First,  recall that the quadrupole moment is evaluated on $\scri^+$ and that $t_{\text{ret}} \hat{=} - \frac{1}{H}\ln (Hr)$.
This simplified the calculation, as we took the late time limit of the quadrupole evaluated at retarded time \emph{before} calculating the transverse-traceless components. 
Second, although in  Sec.~\ref{sec:physicalsetup}, we are interested in the high-frequency limit of $Q_{ab}^{TT}$, the above expressions are true to all order in $\frac{H}{\Omega}$ (assuming $\frac{d R_*}{dt}=0$, not just $\frac{d R_*}{dt}=\ord (\sqrt{\Lambda}t_c)$).
The differential equations mix powers of $\frac{H}{\Omega}$.
This mixing is responsible for the complicated form of $Q_{ab}^{TT}$ as compared to $Q_{ab}$ itself. 
To obtain the high-frequency limit, we truncate the result in the end.
As shown in \eref{eq:diffTTvstt}, taking the high-frequency limit of $Q_{ab}^{TT}$ does \emph{not} reduce $Q_{ab}^{TT}$ to $Q_{ab}^{tt}$.

Given the above recipe, one could in principle also apply this to the radiation field $\mathcal{R}_{ab}$, appearing in the formula for power, to extract its transverse-traceless components. 
This is not needed, however. 
Since the decomposition in \eref{eq:Decomp} is done on a spatial slice, it commutes with time derivatives. 
However, it generically does not commute with spatial derivatives. 
A notable exception to this is the derivative along the `dilation' vector field, $r \del_r$, that plays the role of generating time translations on $\scri^+$ of de~Sitter spacetimes. 
Therefore, $\left(\Lie_T Q_{ab}\right)^{TT}=\Lie_T Q_{ab}^{TT}$ and knowing the transverse-traceless components of $Q_{ab}$, one can easily obtain the transverse-traceless part of the radiation field $\mathcal{R}_{ab}$ by simply calculating the appropriate Lie derivatives of $Q_{ab}^{TT}$.
To obtain the high-frequency limit of $\mathcal{R}_{ab}^{TT}$, we first calculate $\mathcal{R}_{ab}^{TT}$ using this method and truncate the result in the end.


\bibliographystyle{unsrt}

\bibliography{deSitterGWv2}

\end{document}